\documentstyle[12pt,preprint,revtex]{aps}
\begin{document}
\def\overlay#1#2{\setbox0=\hbox{#1}\setbox1=\hbox to \wd0{\hss #2\hss}#1%
\hskip -2\wd0\copy1}
\begin{title}
\begin {center}
{ IMPURITIES IN ``ODD-PAIRING''SUPERCONDUCTORS }
\end{center}
\end{title}
\author{E.\ Z.\ Kuchinskii and M.\ V.\ Sadovskii}
\begin{instit}
Institute for Electrophysics,\ Russian Academy of Sciences,\
Ekaterinburg 620219,\ Russia
\end{instit}
\begin{abstract}
We present the results of theoretical analysis of normal impurities effects in
superconductors with the gap being an odd function of $k-k_{F}$.\ This model
proposed by Mila and Abrahams leads to the possibility of pairing in the
presence of an arbitrarily strong short-range repulsion between electrons and
may beapplied to high-$T_{c}$ oxides.\ However,\ we demonstrate
that normal impurities lead to rather strong suppression of this type of
pairing,\ which is actually stronger than in the case of magnetic impurities
in traditional superconductors.\ Relative stability of high-$T_{c}$ cuprates
to disordering makes this model a rather unlikely candidate for the pairing
mechanism in these systems.
\end{abstract}
\vskip 3.5cm
{\sl Submitted to JETP Letters, March 1993}
\newpage
\vskip 0.5cm
\narrowtext
In a recent paper Mila and Abrahams proposed an interesting model,\
which allows the existence of superconducting pairing even in the case of
infinitely strong point-like repulsion between electrons\cite{1}.\ Naturally
this model is of great interest as a basis for a possible mechanism of
high-temperature superconductivity in metallic oxides.\ The  model is based
upon the demonstration of the existence of nontrivial solution of BCS-like
gap equation:
\begin{eqnarray}
\Delta(\xi)=-N(0)\int\limits_{-\omega_{c}}^{\omega_{c}}d{\xi'}V(\xi,\xi')
\frac{\Delta(\xi')}{2\sqrt{\xi'^{2}+\Delta^{2}(\xi')}} th\frac{\sqrt{\xi'^{2}+
\Delta^{2}(\xi')}}{2T}
\end{eqnarray}
with the gap function $\Delta(\xi)=-\Delta(-\xi)$ (i.e.\ odd in
$k-k_{F}$,\ $\xi=v_{F}(k-k_{F})$) in case of the presence in $V(\xi,\xi')$
of an attractive interaction $-V_{2}(\xi,\xi')<0$ (which is non-zero
for $|\xi|,|\xi'|<\omega_{c}$
and $|\xi-\xi'|<\omega_{c}$) despite the existence of a strong (infinite)
point-like repulsion  $V_{1}(\xi,\xi')=U>0$ (for $|\xi|,|\xi'|<E_{F}$).\
In case of the odd gap function $\Delta(\xi)$ the repulsive interaction
in Eq.\ (1) drops out,\ while the attractive part $V_{2}(\xi,\xi')$ may
produce pairing with unusual properties (gap function is zero at the Fermi
surface,\ which leads to the gapless superconductivity).

If the normal (nonmagnetic) impurities are present the equations for normal
and anomalous Green's functions take the usual form\cite{2},\ which is valid in
case of weak scattering:
\begin{eqnarray}
G(\omega \xi)=-\frac{i\tilde\omega + \xi}{\tilde\omega^{2}+\xi^{2}+
|\tilde\Delta(\xi)|^{2}} \nonumber\\
F(\omega \xi)=\frac{\tilde\Delta^{\star}(\xi)}{\tilde\omega^{2}+\xi^{2}
+|\tilde\Delta(\xi)|^{2}}
\end{eqnarray}
где $\omega=(2n+1)\pi T$,
\begin{eqnarray}
\tilde\omega=\omega-\frac{\gamma}{\pi}\int\limits_{-\infty}^{\infty}d\xi
\frac{\tilde\omega}{\tilde\omega^{2}+\xi^{2}
+|\tilde\Delta(\xi)|^{2}} \nonumber\\
\tilde\Delta(\xi)=\Delta(\xi)
+\frac{\gamma}{\pi}\int\limits_{-\infty}^{\infty}d\xi
\frac{\tilde\Delta(\xi)^{\star}}{\tilde\omega^{2}+\xi^{2}
+|\tilde\Delta(\xi)|^{2}}=\Delta(\xi)
\end{eqnarray}
Here $\gamma=\pi cV^{2}_{0}N(0)$---is the scattering rate due to point-like
impurities with potential $V_{0}$,\ chaotically distributed in space with
concentration $c$.\ The integral term in the second equation vanishes due to
the odd nature of $\Delta(\xi)$ and the gap renormalization is absent.\
This fact explains the strong impurity suppression of the  ``odd'' pairing.\
Note that the similar behavior exists in the case of anisotropic pairing e.g.\
of the  $d$-wave type\cite{3,4}.

The gap equation takes now the following form:
\begin{equation}
\Delta(\xi)=N(0)T\sum_{\omega_{n}}\int\limits_{-\infty}^{\infty}d\xi'
V_{2}(\xi,\xi')
\frac{\Delta^{\star}(\xi')}{\tilde\omega^{2}+\xi'^{2}+|\Delta^{2}(\xi')|^{2}}
\end{equation}
Close to the transition temperature $T_{c}$ Eqs.\ (3) and (4) may be linearized
over $\Delta(\xi)$,\ and after the standard calculations we obtain the
following
linear gap equation,\ which determines $T_{c}$:
\begin{eqnarray}
\Delta(\xi)=N(0)\int\limits_{-\infty}^{\infty}d\xi' V_{2}(\xi,\xi')
\int\limits_{-\infty}^{\infty}\frac{d\omega}{2\pi}\frac{1}{\xi'}
th\left(\frac{\omega + \xi'}{2T}\right)\frac{\gamma}{\omega^2
+\gamma^{2}}\Delta(\xi')
\end{eqnarray}
In the following we shall use the model interaction:
\begin{equation}
V_{2}(\xi,\xi')=\Biggl\{\begin{array}{l}
V[cos \frac{\pi}{2}\frac{\xi-\xi'}{\omega_{c}}+1] \mbox { ; }
|\xi-\xi'|<\omega_{c} \nonumber\\
0 \mbox { for } |\xi|,|\xi'|>\omega_{c};|\xi-\xi'|>\omega_{c}
\end{array}
\end{equation}
The main attractive property of this choice is that it allows the reduction
of the integral gap equation to a simple transcendental equation which can be
easily solved.\ Model potentials used in Ref.\cite{1} do not allow such a
reduction and in most cases have no other serious preferences.\ The main
qualitative results obtained below do not depend on the choice of the model
potential.

The gap function takes now the following form:
\begin{equation}
\Delta(\xi)=\Delta_{0}(T)sin\left(\frac{\pi}{2}\frac{\xi}{\omega_{c}}\right)
\mbox { for } |\xi|<\omega_{c}
\end{equation}
with $\Delta(\xi)=0$ for $|\xi|>\omega_{c}$.\ The  $T_{c}$-equation reduces to:
\begin{eqnarray}
1=N(0)V\int\limits_{0}^{\omega_{c}}\frac{d\xi'}{\xi'}sin^{2}\left(\frac{\pi}{2}
\frac{\xi'}{\omega_{c}}\right)\int\limits_{-\infty}^{\infty}\frac{d\omega}{\pi}
th\left(\frac{\omega+\xi'}{2T_{c}}\right)\frac{\gamma}{\omega^{2}+\gamma^{2}}
\end{eqnarray}
In the ``pure'' limit of ($\gamma\rightarrow 0$) we get the $T_{c}$
dependence on the pairing coupling constant $g=N(0)V$,\ which is shown
in Fig.\ 1.\ Pairing exists for $g>g_{c}=1.213$.\ In Fig.\ 2 we show the
dependence of $T_{c}$ on $\gamma$ for a number of characteristic values
of the pairing constant $g$.\ It is clearly seen that normal impurities
strongly suppress the ``odd'' pairing.\ Superconductivity vanishes for
$\gamma\sim T_{c0}$ and this suppression is even stronger than in case of
magnetic impurities in traditional superconductors\cite{5}.\ This is reflected
in particular by the disappearance of superconductivity region on the
``phase diagramm'' in Fig.\ 2 for $g\rightarrow g_{c}$ and the absence of the
universal behavior which is characteristic for the case of magnetic impurities.

The critical scattering rate  $\gamma_{c}$,\ corresponding to
$T_{c}(\gamma\rightarrow \gamma_{c})\rightarrow 0$,\ is determined,\
according to Eq.\ (5),\ by the following equation:
\begin{equation}
\Delta(\xi)=N(0)\int\limits_{-\infty}^{\infty}d\xi' V_{2}(\xi,\xi')
\frac{1}{\pi\xi'}arctg\left(\frac{\xi'}{\gamma_{c}}\right)\Delta(\xi')
\end{equation}
which for the model interaction of Eq.\ (6) reduces to:
\begin{equation}
1=\frac{2}{\pi}N(0)V\int\limits_{0}^{\omega_{c}}\frac{d\xi'}{\xi'}
sin^{2}\left(\frac{\pi}{2}\frac{\xi'}{\omega_{c}}\right)arctg\left(
\frac{\xi'}{\gamma_{c}}\right)
\end{equation}
It is easily shown that for $g\gg g_{c}$ we have the universal result:
$\gamma_{c}/T_{c0}=4/\pi\approx 1.273$.\ It is also not difficult to see that
this result as well as the dependence of $T_{c}$ on $\gamma$ for $g\gg g_{c}$
do not depend at all on the choice of the model potential $V_{2}(\xi,\xi')$.\
For $g\rightarrow g_{c}$ we always obtain the dependence $\gamma_{c}\sim
(g-g_{c})\rightarrow 0$.\ This behavior is clearly seen in Fig.\ 2.

We have already noted that the model under discussion is attractive as a
possible microscopic approach for the explanation of high-temperature
superconductivity in metallic oxides\cite{1}.\ It is well known that
high-$T_{c}$ state in these systems is very sensitive to the structural
disordering\cite{6}.\
However,\ from the existing experimental data\cite{6} it follows that
superconductivity
in oxides is destroyed close to the metal-insulator transition induced by
disordering i.e.\ for $\gamma\sim E_{F}$,\ but not for
$\gamma\sim T_{c0}\ll E_{F}$.\
This fact makes the model of an ``odd'' pairing rather improbable candidate
for the explanation of high-$T_{c}$ superconductivity in cuprates.

\vskip 0.5cm
The authors are grateful to M.\ A.\ Erkabaev for his help in numerical
calculations.\
This work is supported by the Scientific Council on High Temperature
Superconductivity under the grant $N^{o}$ 90135 of the State Research
Program on Superconductivity.

\newpage
\begin{center}
{\bf Figure captions:}
\end{center}
\vskip 0.5cm
Fig.\ 1.\ $T_{c0}$ dependence on the pairing constant $g=N(0)V$ for the model
interaction of Eq.\ (6).

\vskip 0.5cm
Fig.\ 2.\  $T_{c}$ dependence on the scattering rate $\gamma$
for the different values of pairing constant $g$:

1---$g=1.22$;\ 2---$1.24$;\ 3---$1.30$;\ 4---$1.5$;\ 5---$2.0$;\
6---$5.0$,\ 7---$10.0$.

\end{document}